\documentstyle[11pt,newpasp,twoside,epsf]{article}
\def\edcomment#1{\iffalse\marginpar{\raggedright\sl#1\/}\else\relax\fi}
\reversemarginpar

\begin{document}
\title{Periodicities in ASCA X-ray observations of Non-Magnetic Cataclysmic Variables} 

\author{D.S. Baskill, P.J. Wheatley \& J.P. Osborne} 

\affil{Department of Physics \& Astronomy, University of Leicester, University Road, Leicester LE1 7RH, United Kingdom}

\begin{abstract}

  We have carried out a thorough analysis of non-magnetic cataclysmic variables observed with {\it ASCA}, and have discovered possible periods in three targets:  WW~Cet, EI~UMa and IX~Vel. 
These may be caused by the rotation of the white dwarf at the centre of the disk, or by the progression of a density wave in the inner accretion disk.
\end{abstract}

%\section{Introduction}

\vspace{-6 mm}
\section*{Periodic X-ray Modulations in Non-Magnetic CVs}

%Non-magnetic cataclysmic variables are binary systems consisting of a white dwarf accreting from a main sequence secondary star, via a disc.  These systems exhibit periods of optical outburst and quiescence due to variations in the viscosity of the disk, which lead to changes in the mass transfer rate through the disk, and hence luminosity. 

 X-ray observations of non-magnetic cataclysmic variables probe the inner region of the accretion disc, where material slows from a Keplerian velocity in the disk to accrete onto the white dwarf surface.  The Japanese X-ray satellite ASCA made 26 observations of 21 non-magnetic CVs, including 14 dwarf novae and 7 nova likes.

%\section{Periodic Modulations}

  X-ray modulations have possibly been detected in 3 of the targets (WW~Cet, EI~UMa and IX~Vel).  Our highest confidence periods are a 9.8 minute periodicity detected in WW~Cet, and a 12.4 minute period detected in EI~UMa.  Both these periods are well separated from the red noise that tends to dominate the low frequency end of the power spectra.  The third possible period is a 25.6 minute period detected in IX~Vel, but this is well within the red noise range and so it is difficult to access its significance (see figure~1).

  Since the X-rays originate from close to the white dwarf, these periods may be due to the rotation of a weakly magnetised white dwarf at the centre of the disk.  Although no rotational velocity measurements have been made of our 3 targets, Sion et al. (1994) used {\it HST} observations to measure the rotational velocity of the white dwarf in U~Gem to be 50-100\,km\,s$^{-1}$.  By assuming the radius of the white dwarf is 10$^9$\,cm, this implies that the rotational period of the white dwarf is 10-20\,minutes, similar to the observed periods presented here (see Sion 1999 for review of white dwarf rotation in CVs).  These values are far from breakup rotation rates theory predicts (eg. Livio \& Pringle, 1998).
%  A boundary layer should liberate half of the total gravitational energy of accreting matter.  However, boundary 1layers are observed to be less luminous than this, suggesting that the white dwarf maybe rapidally rotating at close to breakup.

 Another possibility is that the modulations arise from a prograde travelling wave in the inner disk, which obscures and/or reprocesses radiation from the central region (Warner \& Woudt, 2001).

\begin{figure}[t]
\plotone{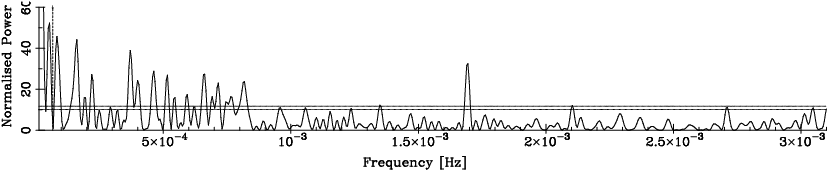}
\plotone{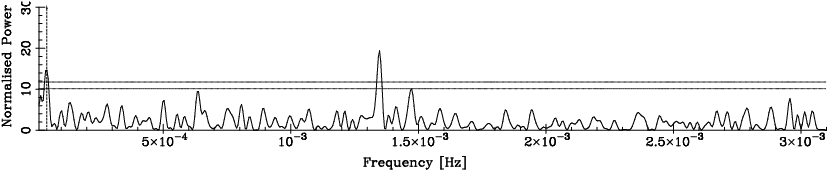}
\plotone{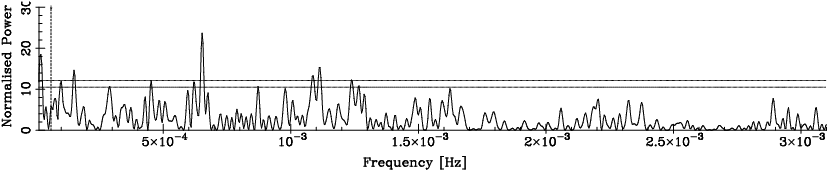}
\vspace{-5 mm}
\caption{A 9.8 minute period is detected in WW~Cet (top) and a 12.4 minute period is detected in EI UMa (centre).  In IX~Vel (bottom) a possible 25.6 minute period is detected, although the significance of this peak difficult to quantify since it appears within the range of red noise.  The horizontal lines mark the 95\% and 99\% confidence levels, and the vertical dotted lines show the orbital period.}
\end{figure}

\begin{figure}[!h]
\plotone{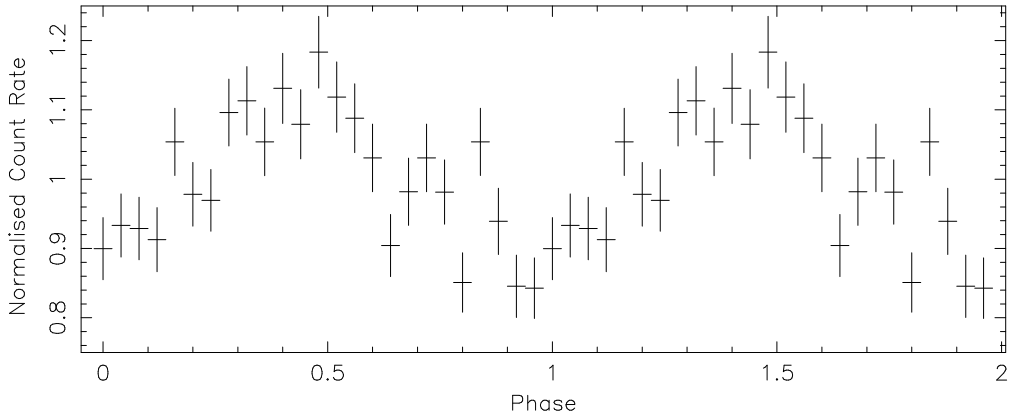}
\vspace{-5 mm}
\caption{The folded {\it ASCA} X-ray light-curve of the 9.8 minute modulation in WW~Cet.}
\vspace{-5 mm}
\end{figure}

\vspace{+3 mm}

\end{document}